\documentclass[preprint, 10pt, 5p, twocolumn]{elsarticle}
\usepackage{amssymb}
\usepackage{amsmath}
\usepackage{float}
\usepackage{hyperref}
\usepackage{lineno}
\usepackage{bm}
\usepackage{bbold}
\usepackage{subcaption}
\usepackage[percent]{overpic}
\journal{Applied Radiation and Isotopes}

\begin{document}
\begin{frontmatter}
 \title{On the Measurability of True Coincidence Summing in the GRIFFIN Spectrometer.}
\author{Liam L. Schmidt \corref{cor1}\fnref{label2}}
\ead{lschmi04@uoguelph.ca}

\affiliation[label2]{organization={Department of Physics},
         addressline={ University of Guelph},
            city={Guelph},
            postcode={N1G 2W1},
                 country={ON, Canada}}

\begin{abstract}
The measurement of gamma-rays from decaying nuclei allow for the investigation into nuclear structure. True coincidence summing occurs when two gamma-rays from a single decay get detected in a single detector and their energies sum together to give false peaks in the energy spectrum. Corrections to this summing effect are crucial for an accurate determination of nuclear decay events. The summing correction formalism of Semkow et.al \cite{SEMKOW1990437} is generalized here into the \textit{Semkow matrix formalism} and is extended into a multiplicity expansion. This formalism is used to calculate the matrix probabilities for 180-degree coincidence events as a method for correcting coincidence summing; in doing so, the deviation between the full correction and this 180-degree correction is shown as a function of multiplicity. This formalism is extended to the \textit{partitioned matrix formalism}, to calculate probabilities for \textit{gated} gamma rays; where two gamma-rays of interest are taken in multi-detector coincidence. The summing correction probabilities for gated gamma-rays are provided and the deviation is shown in a manner similar to that of the singles. Terms such as measurability, event equivalence, and ontic and epistemic events are defined. It is shown that within these definitions, coincidence summing is not sufficiently measurable, or rather, its sufficient measurability is statistically bounded by the deviations derived.
\end{abstract}

\begin{keyword}
True coincidence summing \sep Gamma-ray spectroscopy \sep Summing corrections \sep $\beta$ decay
\end{keyword}

\end{frontmatter}
\section{Introduction}
\begin{figure}[H]
    \centering
    \includegraphics[width=0.9\linewidth]{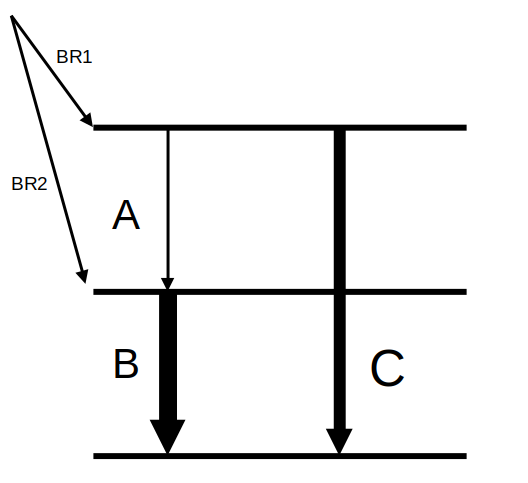}
    \caption{Basic decay scheme for coincidence summing.}
    \label{fig:fig1}
\end{figure}

In gamma ray spectroscopy, gamma rays emitted from decaying nuclei are measured and the corresponding counts of the photopeaks produced by the detection of the gamma rays  are used to determine properties of the nuclei in question. The validity of the results from such a measurement rests on the reliability of the observation being a description of the true event. However, the observation is never an exact description of the event because of unavoidable phenomena that occur in gamma ray spectroscopy. Reliability is recovered by applying corrections to the observations, such as detector efficiency, energy calibrations, pile-up effects, etc. The phenomenon analyzed here is that of coincidence summing, where multiple gamma-rays are deposited into a detector in a small enough time window that their energies get summed together and is observed as a single gamma ray with energy equal to their summed energy \cite{SEMKOW1990437,osti_4488528,MCCALLUM1975189}. This work solely focuses on \textit{true coincidence summing} where only gamma-rays from a single decay are considered. This is distinguished from random coincidence summing, where gamma-rays from different decaying nuclei can sum together in a detector. From here on when stating coincidence summing, only true coincidence summing is considered and is  called coincidence summing. A simple example of this is shown in Fig.~\ref{fig:fig1}, where gamma A and gamma B could sum together to give an apparent gamma C. The term \textit{summing-in} refers to two gamma-rays summing together to add counts to a gamma peak that corresponds to an emitted gamma ray. In this case, counts are lost from peaks A and B and added to C. The term $summing-out$ refers to the counts lost in a photo peak due to full energy gamma-rays summing with other gamma-rays of any energy. The analysis into coincidence summing done here is directed toward calculating coincidence summing effects in the GRIFFIN spectrometer at TRIUMF in Vancouver, Canada \cite{Svensson2014,Garnsworthy_2019}.

\subsection{History of Summing Corrections}
A theoretical account of the effects of coincidence summing and a calculation for their correction was first produced by Andreev \textit{et al.} (1972) \cite{osti_4488528}; they provide a recursive relation for the summing corrections which was reproduced in McCallum \& Coote (1975) \cite{MCCALLUM1975189}. In the treatment of coincidence summing in gamma ray spectroscopy, attention may be given solely to the prompt gamma-rays of the decay, or, in addition, to what may be called \textit{auxiliary radiation}. The auxiliary radiation includes annihilation radiation from positrons, X-rays from internal conversion, bremsstrahlung from beta particles, and all other radiation that may be present in the event of a decaying nucleus as a result of the decay itself, neglecting external contributors. An accurate representation of coincidence summing must include the effects of auxiliary radiation; and since the work done by \cite{osti_4488528, MCCALLUM1975189}, continuous attention has been paid to such effects. The equations in \cite{MCCALLUM1975189} provide the summing corrections for a general decay scheme where summing with auxiliary radiation is not included, but the authors mention the importance. The work by Debertin \& Schotzig (1979)\cite{DEBERTIN1979471} takes the formulae of \cite{osti_4488528,MCCALLUM1975189} and develops the program \textit{KORSUM }to calculate the summing corrections for a given decay that includes summing with X-rays from internal conversion, though no formalism was explicitly shown for dealing with X-ray summing. In the measurement of peak intensities, Gehrke \textit{et al.} (1977)\cite{GEHRKE1977405} include the use of summing corrections where X-rays from internal conversion are included. In Schima \& Hoppes (1983) \cite{SCHIMA19831109}, a table is given that provides summing coefficients for common radionuclides using the formulae of \cite{osti_4488528, MCCALLUM1975189}. In 1983 Morel \textit{et al.} \cite{MOREL19831115} provided similar expressions for the summing probabilities using product notation which Tomarchio \& Rizzo \cite{TOMARCHIO2011318, RIZZO2010555} used to provide corrections in gamma ray spectroscopy in high-purity germanium (HPGe) detectors. The work extended in this paper is that of Semkow et.al \cite{SEMKOW1990437}, where the authors take the recursive relations of \cite{osti_4488528,MCCALLUM1975189} and develop a matrix formalism. In their derivation, no auxiliary radiation is included in the summing but internal conversion probabilities are included in the efficiencies. In the same year, Richardson \& Sallee \cite{RICHARDSON1990344} provided an analysis on the coincidence summing corrections for positron emitters where summing with the 511keV gamma-rays emitted by positron annihilation contribute to the coincidence summing.  In 1992, Korun \textit{et al.} \cite{KORUN1993478} extended the matrix formalism of Semkow to include X-rays from internal conversion by including virtual branches in the decay scheme. After this period, the majority of the literature that addresses coincidence summing uses the work mentioned previously in computational analyzes using Monte Carlo simulations. Several programs have been developed that provide corrections for certain aspects of coincidence summing for gamma ray spectroscopy such as GESPECOR by Arnold \& Sima\cite{SIMA200051,ARNOLD2000725, ARNOLD2004167, ARNOLD20061297} and ETNA \cite{PITON2000791}; other computational analyzes are done in \cite{HAASE1993483,KANISCH20091952, DZIRI20121141, GIUBRONE2016114}. In 2007, Jutier \textit{et al.} \cite{JUTIER20071344} provided a new synthetic formalism that deals with summing corrections with complete treatment of auxiliary radiation in a manner different from the matrix formalism of Semkow. They went on to show, through the neglect of auxiliary radiation, that their formalism retrieves the probabilities present in \cite{osti_4488528, MCCALLUM1975189}. The formalism they provide, however, is offered in a manner less explicit than previous work as the authors state the work was implemented into a program called \textit{Coincal} where they provide corrections for the decay of $^{152}$Eu. In addition, much has been done on coincidence summing from extended sources \cite{DECOMBAZ1992152, VIDMAR2006543,VIDMAR2007243, ABBAS2007554} which is, however, not of concern in this present work. There has also been work on empirical methods for summing corrections, e.g., in \cite{QUINTANA1995961, RAMOSLERATE1997202},  where differences in detector geometries are compared.
More recent work on coincidence summing corrections can be seen in \cite{Badawi1, Badawi2, Badawi3}, where the authors provide a powerful and succinct mathematical formulation on coincidence summing, highlighting the importance of understanding this phenomena for the calibration of detectors with different sources.
\par
We categorize the summing corrections present in the literature under the guises of being either theoretical, computational, or empirical; where theoretical and computational may be grouped together as \textit{a priori}, i.e., \textit{not} deduced directly from observation or experience.

\subsection{The Ontic-Epistemic Division of Events}
The following terms are defined to explicate notions of detection. An event in its fullness that exists external to our observation is an \textit{ontic event}. It is the ontic events one wishes to understand in order to arrive at properties of the nuclei in question. In attempts to measure such ontic events, what is observed are \textit{epistemic events}. These are events that correspond to ontic events but have been corrupted or have lost information due to limitations of measurability. It should be noted here that ontic events are categorical; they are generalized occurrences with a qualitative description. In contrast, epistemic events are singular; they are singular happenings in space and time.
\par
In an experiment, one gathers a set of epistemic events and attempts to map these to the set of ontic events, in hopes that the properties of the latter may afford insight into the object of the experiment. In the case of gamma-ray spectroscopy, an ontic event would be a certain decay of the nucleus where the resultant radiation is emitted in a particular direction. In an experiment, such an event may yield numerous epistemic events based on the measuring apparatus and the particularities of the experiment. A corresponding epistemic event may look like a set of data that contains values for energy, position, and time. Those values, of course, may not directly equate to the energy, position, and time of the radiation emitted by the nucleus. 
\par 
Let us now further define equivalences between events and notions of measurability. A \textit{complete equivalence} exists between ontic and epistemic events if the epistemic event is a true description of the ontic event, i.e., no corruption occurs in the measurement. A \textit{hidden ontic event} is an ontic event whose measurement yields an epistemic event that is not distinct, i.e., no complete equivalence can exist.  An ontic event is \textit{measurable} if it can yield a corresponding epistemic event that exists and is distinct. A \textit{probabilistic equivalence} exists between ontic events if there is an equality between their probabilities of occurrence. A hidden ontic event is \textit{sufficiently measurable} if it is probabilistically equivalent to a distinct measurable ontic event and is \textit{theoretically recoverable} if it can be accounted theoretically, i.e., by the use of a correction factor.
\par
 Putting context to these definitions, coincidence summing--or rather, the zero-degree emission of two photons--is defined as a hidden ontic event, i.e., a phenomena that exist but its observation is not distinct. In contrast, the 90-degree emission of gamma A and B may yield an epistemic event that may be completely equivalent to the ontic event. The summing-in, however, of gamma A and B giving an epistemic event of full energy C can not offer a complete equivalence because it can not be mapped to a distinct ontic event. The zero degree emission of gamma A and gamma B and the emission of gamma C may yield the same epistemic events. To correct for coincidence summing events, one either attempts to theoretically recover them or sufficiently measure them. Theoretically recovering the lost counts may look like multiplying the total peak area by some factor that attempts to restore the true value. 
 
 A common method used to correct for coincidence summing is the \textit{180-degree coincidence method} \cite{Garnsworthy_2019}. In this manner one sufficiently measures coincidence summing by stating an equivalence between two gamma-rays hitting the same detector and two gamma-rays being in 180-degree coincidence, i.e., with both events occurring with the same probability one may measure the 180-degree coincidence events and use these counts to sufficiently measure the coincidence summing counts. The equivalence between these distinct events is an equality of probabilities. Thus, when realized experimentally, the measurability is \textit{statistically sufficient} to the degree defined by the experiment. This work sets out to show that the equivalence between 180-degree coincidence events and summing coincidence is not \textit{completely sufficient} due to the inseparability of the 180-degree coincident events and the summed events; an inseparability that increases with event multiplicity, i.e., the number of gamma rays emitted in a decay. Multi-detector coincidence events often used in gamma-ray spectroscopy to \textit{gate} on certain decay branches and the coincidence summing that occurs there are also considered and a matrix formalism for these probabilities are provided.

\section{Coincidence Summing}
\subsection{The GRIFFIN Spectrometer}
\begin{figure}[]
    \centering
    \includegraphics[width=0.9\linewidth]{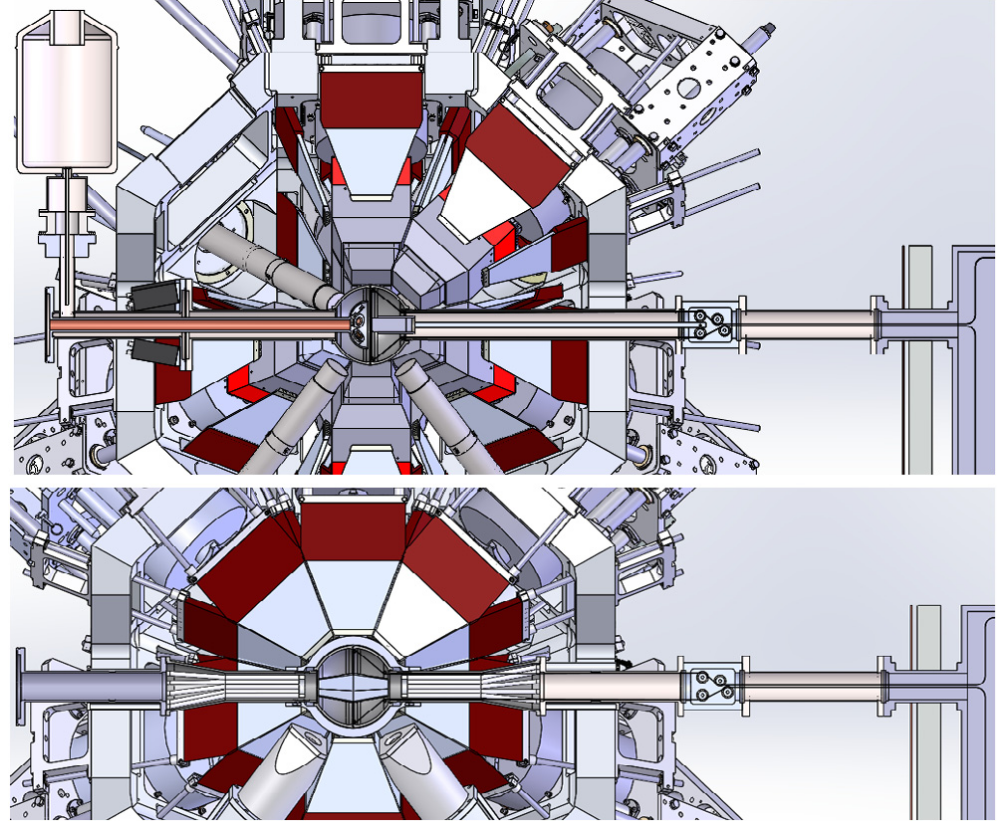}
    \caption{Two possible configurations for the GRIFFIN spectrometer \cite{Garnsworthy_2019}.}
    \label{fig:fig3}
\end{figure}

\begin{figure}[H]
    \centering
    \includegraphics[width=1.0\linewidth]{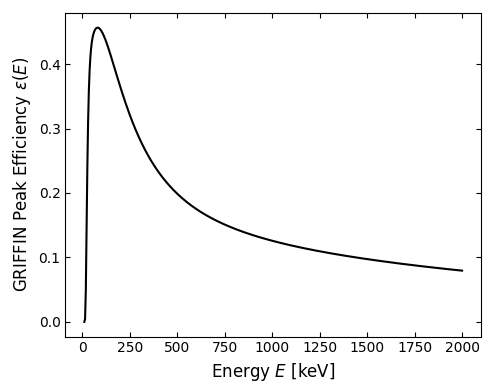}
    \caption{Simulated peak efficiencies for GRIFFIN \cite{Mills2015}.}
    \label{fig:fig2}
\end{figure}

\begin{figure}[]
    \centering
    \includegraphics[width=0.7\linewidth]{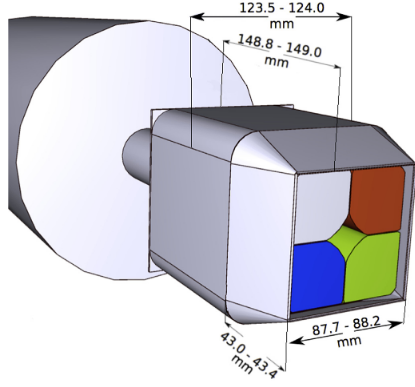}
    \caption{Model of the GRIFFIN HPGe clover detector \cite{RIZWAN2016126}.}
    \label{fig:fig4}
\end{figure}

This work focuses on the coincidence summing that occurs in the GRIFFIN spectrometer \cite{Svensson2014,Garnsworthy_2019,GARNSWORTHY201785}--though may apply to similar detector arrangements. The GRIFFIN spectrometer is a multi-detector array of HPGe clover detectors as shown in Fig~\ref{fig:fig3}. Each of the 16 clovers composing the entire array is made of 4 n-type HPGe crystals as shown in Fig~/\ref{fig:fig4}. Each of the 16 detectors makes up a square face of the 18 sides of a rhombicuboctahedron; the remaining two squares being used for the in-line radioactive beam which deposits nuclei onto a moving tape system that removes long-lived activity from the chamber at the end of each measurement cycle \cite{Garnsworthy_2019}. A plot of its simulated peak efficiency is shown in Fig.~\ref{fig:fig2}.\footnote{11 cm, Single crystals, no shields, US+DS SCEPTAR, no delrin} 
The considerable volume of the germanium crystals in each clover contributes to its high efficiency. However, as is well known, the coincidence summing effects present in a detector increases with solid angle. In GRIFFIN, coincidence summing can result in a shift of $1\%-10\%$ on the measured efficiency for different energies.
The GRIFFIN collaboration attends to summing effects via a 180-degree correction \cite{Garnsworthy_2019}. In \cite{Garnsworthy_2019}, the 180-degree correction method is used to calculate the summing-out for peaks in the decay of $^{152}$Eu and report a summing-out percentage of approximately $3\%-6\%$. It is stated in \cite{Garnsworthy_2019}, that this correction gives the coincidence summing \textit{within statistics}. The deviations calculated herein are intended to probe the limits of the statistical precision of this correction.

\subsection{Semkow Matrix Formalism}
\begin{figure}[]
    \centering
    \includegraphics[width=0.9\linewidth]{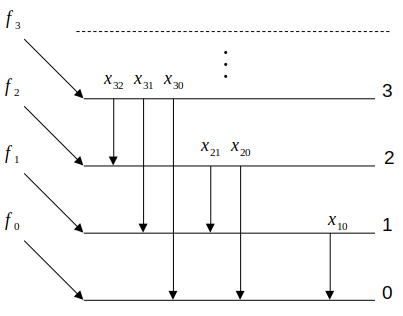}
    \caption{General decay scheme diagram.\cite{SEMKOW1990437}}
    \label{fig:genDecay}
\end{figure}

The general matrix formalism for coincidence summing given by Semkow \textit{et al.}, with the general decay is shown in Fig.~\ref{fig:genDecay}, is as follows. It is assumed that some parent nucleus decays through any possible branch $f_i$ to a daughter nucleus with $n$ levels. Each level has a probability of decay to each branch below it with transition probability $x_{ij}$,  where the transition is from level $i$ to level $j$. In Fig.~\ref{fig:genDecay}, only 3 levels are shown, but in the following derivation $n$ levels are considered. It should be noted that the level spacing in Fig.~\ref{fig:genDecay} is not representative of gamma energy that is typical in decay diagrams, it is merely taken as a general decay scheme.
Each transition probability is put into a lower triangular \textit{transition matrix}
\begin{equation}
    x = \begin{pmatrix}
        0 &\\
        x_{10} & 0 \\
        x_{20} & x_{21} & 0\\
        \vdots & \vdots & \vdots \\
        x_{n0} &x_{n1} & x_{n2} & \cdots& x_{n n-1} & 0
    \end{pmatrix}.
\end{equation}
The branching ratios can be put into a \textit{branching vector} $f$,
\begin{equation}
    f = (f_0, f_1, f_2, \cdots, f_n).
\end{equation}
Each transition is associated with some gamma-ray of interest (GOI) $\gamma_{ij}$ with energy $E_{ij}$. The elements of the branching vector and the transition matrix satisfy the following conditions:
\begin{equation}
    \sum_{i=0}^n f_i = 1,
\end{equation}
\begin{equation}
    \sum_{j=0}^{i-1} x_{ij} = 1,  \hspace{20pt} i = 1,2,...n  .
\end{equation}
The above steps follow directly from \cite{SEMKOW1990437}. At this point, the authors in \cite{SEMKOW1990437} go on to define experimental probabilities using the transition probabilities and efficiencies. In a similar manner, Eqs.~\eqref{eqn:a_matrix}-~\eqref{eqn:l_matrix} are given, where $\epsilon_{ij}^p$ gives the peak efficiency and $\epsilon_{ij}^T$ gives the total efficiency.\footnote{In \cite{SEMKOW1990437}, internal conversion coefficients are given--this step is ignored here but could easily be appended to the formalism. In general, any GOI matrix may be allowed into the $\Gamma(Q,\Lambda_\nu)$ function defined herein.} 
\begin{equation}
    a_{ij} = x_{ij}\epsilon_{ij}^p,
    \label{eqn:a_matrix}
\end{equation}
\begin{equation}
    \omega_{ij} = \frac{x_{ij}\epsilon_{ij}^T}{N},
    \label{eqn:w_matrix}
\end{equation}
\begin{equation}
    {\lambda_{(\nu)}}_{ij} = x_{ij}\left(1-\nu\frac{\epsilon_{ij}^T}{N}\right).
        \label{eqn:l_matrix}
\end{equation}
The above matrices differ from Semkow \textit{et al.} here by generalizing the summing-out probabilities to $\lambda_\nu$ where $\nu=1$ gives the $b$ matrix given by Semkow et al. An experimental set up like GRIFFIN is also considered here, consisting of $N$ detectors. The other matrix defined here, $\omega$, assists in the coincidence probabilities to be calculated in the proceeding sections. The above matrices are power expanded to create cascade terms where the $N$ dependence gives the probability of hitting the same detector--ignoring any angular correlations:
\begin{equation}
    A = \sum_{k=1}^n \frac{a^k}{N^{k-1}},
\end{equation}
\begin{equation}
    \Omega = \sum_{k=1}^n \omega^k,
\end{equation}
\begin{equation}
    \Lambda_\nu = \mathbb{1} + \sum_{k=1}^n \lambda_\nu^k.
\end{equation}
At this point, Semkow et al. goes on to take the matrices $A$ and $\Lambda$ (with $\nu=1$) into a formalism that creates two diagonal matrices using $\Lambda_\nu$ to account for summing-out from gamma-rays above and below the GOI. This is generalized to the \textit{Semkow matrix formalism} (SMF) and given the following function:\footnote{For the purpose of this paper we drop the rate dependence R that is contained in the S matrix given by Semkow}
\begin{equation}
    \Gamma(Q,\Lambda_\nu) = \mathcal{D}([f\Lambda_\nu]_i) Q \mathcal{D}([\Lambda_{\nu}]_{i0}),
\end{equation}
where $\mathcal{D}()=diag()$ indicates the diagonal matrix with elements of those contained in the brackets. This generalization is motivated by the pursuit to calculate the probabilities for different phenomena, not just the summing-out of gamma-rays, but \textit{selective coincidence phenomena} that can be contained in the $\Lambda_\nu$ matrix by choice of $\nu$, where summing is considered a type of coincidence phenomena, \textit{viz.}, the zero degree coincidence.\footnote{By selective coincidence phenomena here, it is meant that the $\nu$ selects the number of detectors involved in the coincidence. A coincidence of $\nu=1$ gives same detector coincidence and thus summing. With $\nu=2$, standard two detector coincidence can be selected. This is allowed under the isotropic emission approximation.} The input matrix $Q$ for the SMF contains probabilities for the GOI where summing-in into the GOI can occur, given by a power expansion like that of equations 8-10. For singles spectra, $Q$ will take the form of $A$. However, this is, in general, not always the case. When looking at gated probabilities this will change. The \textit{singles decay matrix}, $S$, which gives the probabilities of observing a GOI accounting for summing-in and summing-out is thus given by,
\begin{equation}
    S = \Gamma(A,\Lambda_1).
\end{equation}
As an example, to make the output of the SMF more clear, below is the singles decay matrix for a decay scheme with $n=2$:
\begin{equation}
    S = \begin{pmatrix}
        0 &0&0\\
        (f_1 + f_2b_{21})a_{10} & 0 & 0\\
        f_2\left(a_{20} + \frac{a_{21}a_{10}}{N}\right) & b_{10}f_2a_{21} & 0\\
         \end{pmatrix}.
\end{equation}
Looking at $\gamma_{10}$, its probability of detection comes from the feeding from branch one and the feeding from branch two weighted by the probability that it does not sum out with $\gamma_{21}$. Looking at $\gamma_{20}$, it is only fed from branch two but can come from the true $\gamma_{20}$ or the summing of $\gamma_{21}$ and $\gamma_{10}$. 

\subsection{Multiplicity Events}
The SMF is now extended into an expansion of \textit{multiplicity terms}.
The gamma matrix can be separated into terms of equal branching, where for ease of notation $\lambda_1 \equiv b$ with $\Lambda_1 \equiv B$, momentarily restoring the notation of Semkow.
\begin{equation}
    \Gamma(A,B) = \sum_{\mu=1}^ n f_\mu\psi ^{(\mu)} (A,B),
\end{equation}
with,
\begin{equation}
    \psi_{ij}^{(\mu)} (A,B) = A_{ij}B_{\mu i}B_{j0},
\end{equation}
where $\mu$ indicates the branch and thus the maximum multiplicity. Furthermore, $\psi^{(\mu)}$ can be written in terms of a summation over all event multiplicities up to the maximum $\mu$ allowed by the branch. Each term $\phi^{(\mu m)}$ contains elements composed of products of $m$ probabilities either coming from matrix $A$ or $B$:
\begin{equation}
    \psi^{(\mu)} (A,B) = \sum_{m=1}^\mu \phi^{(\mu m)} (A,B).
\end{equation}
In the proceeding equations where multiplicity terms are shown, the exponent notion indicates matrix multiplication with the indices taken of the power; that is, $a^k_{ij} :=[a^k]_{ij}$. As such, $a^0$ indicates the identity matrix. The form of $\phi_{ij}^{(\mu m)}$ is as follows:
\begin{equation}
\begin{split}
     \psi_{ij}^{(\mu)} (A,B) &= A_{ij}B_{\mu i}B_{j0}\\
     &= \left(\sum_{l=1}^\mu \frac{a_{ij}^l}{N^{l-1}}\right) \left(\delta_{\mu i} +\sum_{p=1}^\mu b_{\mu i}^p \right)\left(\delta_{j 0} + \sum_{q=1}^\mu b_{j0}^q\right)\\
     &=\sum_{l=1}^\mu \frac{a_{ij}^l}{N^{l-1}}\sum_{p=0}^\mu b_{\mu i}^p \sum_{q=0}^\mu b_{j0}^q \\
     &=\sum_{l=1}^\mu \sum_{p=0}^\mu\sum_{q=0}^\mu \frac{a_{ij}^lb_{\mu i}^p b_{j0}^q }{N^{l-1}}.\\
\end{split}
\label{eqn:eqn17}
\end{equation}
\par
For the terms in the above sum, the multiplicity of the terms, i.e., the sum of the exponents $m=l+p+q$, is bounded by the branch number $\mu$. Therefore, the only non-zero terms in the above sum are in this bound and the sums need not all be written to the $\mu^{th}$ term. This sum can now be reordered in terms of the multiplicity. The general term of Eq.~\eqref{eqn:eqn17} can be taken as the general term of a set:
\begin{equation}
   \mathcal{U} =  \left\{\frac{a_{ij}^lb_{\mu i}^p b_{j0}^q }{N^{l-1}}\mid m=l+p+q, 1\leq m \leq \mu\right\}.
   \label{eqn:eqn18}
\end{equation}
We establish a total order $\preceq_m$ on $\mathcal{U}$ which orders the elements of the set based on their multiplicity $m$. Two elements $u^{lpq}_1$ and $u^{lpq}_2$ in the set then satisfies $u^{l_1p_1q_1}_1\preceq_m u^{l_2p_2q_2}_2$ \textit{iff} $m_1\leq m_2$. The set $\mathcal{U}$ can now be divided into subsets of equal $m$ and each ordered by the relation $\preceq_l$ which orders the elements of $\mathcal{U}_m$ by $u^{l_1p_1q_1}_1\preceq_l u^{l_2p_2q_2}_2$ \textit{iff} $l_1\leq l_2$. The sets $\mathcal{U}_m$ are defined for $0\leq l \leq m$. This process is continued for all elements in $\mathcal{U}_m$ of equal $l$  defining subsets $\mathcal{U}_{ml}$ ordered by $\preceq_p$ which orders the elements of $\mathcal{U}_{ml}$ by $u^{l_1p_1q_1}_1\preceq_p u^{l_2p_2q_2}_2$ \textit{iff} $p_1\leq p_2$. The subsets $\mathcal{U}_{ml}$ are defined for $0\leq p \leq m-l$. Finally, for all elements in $\mathcal{U}_{ml}$ of equal $p$ giving the subsets $\mathcal{U}_{mlp}$, $q$ is restricted by the multiplicity condition $q=m-l-p$. At this point, the set $\mathcal{U}$ has been fully ordered by ordering the subsets $\mathcal{U}_m \subset \mathcal{U}$  by increasing $m$. Each subset $\mathcal{U}_{ml} \subset \mathcal{U}_m$ is ordered by increasing $l<m$, and finally each subset $\mathcal{U}_{mlp} \subset \mathcal{U}_{ml}$ is ordered by increasing $p<m-l$ leaving $q=m-l-p$ for the single element of each set $\mathcal{U}_{mlp} $. This ordering is now used to sum over nested subsets which will give a summation equal to that of equation 17 but defined and ordered by the multiplicity $m$ of the terms:
\begin{equation}
    \psi^{(\mu)} (A,B) = \sum_{m=1}^\mu  \sum_{l=1}^m \sum_{p=0}^{m-l}\frac{a_{ij}^l b_{\mu i}^pb_{j0}^{m-l-p}}{N^{l-1}}.
    \label{eqn:eqn19}
\end{equation}
With this, the elements $\phi_{ij}^{(\mu m)}$ initially defined in equation 16 can be given:
\begin{equation}
    \phi_{ij}^{(\mu m)} (A,B) = \sum_{l=1}^m \sum_{p=0}^{m-l}\frac{a_{ij}^l b_{\mu i}^pb_{j0}^{m-l-p}}{N^{l-1}}.
    \label{eqn:eqn20}
\end{equation}
\par 
Therefore, with a decay of $n$ branch levels, $\phi_{ij}^{(\mu m)}$ contains the probability for detecting a gamma of energy $E_{ij}$ that came from an event of multiplicity $m$ from branch $\mu$. Note, however, that it does not mean the probability of detecting $\gamma_{ij}$ but only a hit with energy $E_{ij}$ as the hit can come from summing-in. \textit{Ontic multiplicity} and \textit{epistemic multiplicity} are defined, where the former refers to the number of emitted gamma-rays in the decay, and the later refers to the number of detected gamma-rays. The terms in the $\phi^{(\mu m)}$ matrices all refer to events with the same ontic multiplicity but may differ in epistemic multiplicity. The script $m$ is thus ontic multiplicity with a maximum value of $\mu$. The index $l$ gives the summing-in degree for the GOI such that $l=2$ gives the coincidence summing of two gamma-rays. The index p sums over the number of gamma-rays that contribute to summing-out above the GOI, while $m-l-p$ gives the number of gamma-rays below the GOI. Separating out the decay matrix in this manner allows us to select out events of desired multiplicity; a parameter of which determines the likelihood of summing. As an example from our $n=2$ case:
\begin{eqnarray}
    \phi^{(2 2)} (A,B) &=&    \begin{pmatrix}
    0&0&0\\
    b_{21}a_{10} &0&0\\
    \frac{a_{21}a_{10}}{N} &b_{10}a_{21} & 0
    \end{pmatrix},\\
    \phi^{(2 1)} (A,B) &=&    \begin{pmatrix}
    0&0&0\\
    0&0&0\\
    a_{20}&0& 0
    \end{pmatrix},\\
    \phi^{(1 1)} (A,B) &=&    \begin{pmatrix}
    0&0&0\\
    a_{10}&0&0\\
    0&0& 0
    \end{pmatrix}.\\\nonumber
\end{eqnarray}
\par
Note, that the ontic event of $\phi_{20}^{(22)}$ has an epistemic multiplicity of one while $\phi_{10}^{(22)}$ \textit{may} have an epistemic multiplicity of two; the probability does not specify both emitted gamma-rays are detected, just that $\gamma_{21}$ does not sum at least partially with $\gamma_{10}$ that is detected at full energy. The goal of experiments is to use a collection of epistemic events; i.e, observations, to determine properties of purposed ontic events. The mapping however, is not always exact. In the next section, probabilities of summing corrections are compared to the summing probabilities defined above and the deviation between the correction and the true probabilities as a function of ontic multiplicity is calculated.

\section{Summing Corrections}
\subsection{Summing-out}

The summing correction technique analyzed here is the \textit{180 degree coincidence method}, which attempts to sufficiently measure the coincidence summing within statistics \cite{Garnsworthy_2019}. The matrices, $\omega$ and $\lambda_2$, define the probabilities for $\gamma_{ij}$ hitting the 180 degree detector and not hitting both detectors in the 180-degree coincidence, respectively. For the 180 degree coincidence events, at least one of the extra gamma-rays that are not the GOI is required to be in 180 degree coincidence. When dealing with an event with $k$ gamma-rays, all possible 180 degree coincidence events need to be considered. For this reason, a sum over all possible permutations of the extra gamma-rays is needed, i.e., when $1,2,3,...k$ gamma-rays are in the 180-degree detector; doing so involves a sum of nested anti commutators, \textit{viz.} the non-commutative form of the binomial expansion. In a toy model with $n$ indistinguishable gamma-rays all with the same probability $x$ of being detected, the binomial expansion would suffice to provide the probability of the 180-degree coincidence event. In the general case where different branching occurs and each gamma-ray has their respective probability of detection, the \textit{n-choose-k} function of the binomial expansion essentially turns into a sum of different permutations of labels. With Eqs.~\eqref{eqn:w_matrix} and ~\eqref{eqn:l_matrix}, different matrices can be built where $q$ gamma-rays are to be in the 180 degree detector. For each scenario, all permutations of gamma-rays that give us the $q$ gamma-rays in the 180-degree detector must be considered. To do so, powers of $b$ and $b = \omega + \lambda$ are considered, where $\lambda \equiv \lambda_2$ (again, the notation is dropped in hopes of increasing the readability of the multiplicity expansion):

\begin{equation}
    \begin{split}
        \sum_{k=1}^nb^k &= \sum_{k=1}^n(\omega + \lambda)^k \\
         \sum_{k=1}^nb^k& =\sum_{k=1}^n\sum^k_{q=0} \pi(\lambda^{k-q},\omega^q),
    \end{split}
\end{equation}
where $\pi(\lambda^{k-q},\omega^q)$ denotes the \textit{sum} of all possible multiplicative permutations of $\lambda^{k-q}$ and $\omega^q$. At the ends of the sum where either $q=0$ or $k=q$,  $\Lambda$ and $\Omega$ can be extracted.
\begin{equation}
    \begin{split}
         \sum_{k=1}^nb^k& =\sum_{k=1}^n\sum^k_{q=0} \pi(\lambda^{k-q},\omega^q)\\
         B - \Lambda  & = \Omega + \sum_{m=1}^{n-1}\sum_{q=1}^{n-m}\pi(\lambda^{k-q},\omega^q) \\
         B - \Lambda & = \Omega + \sum_{m=1}^{n-1}
         \Pi_m,
    \end{split}
    \label{eqn:eqn25}
\end{equation}
where
\begin{equation}
     \Pi_m = \sum_{q=1}^{n-m}\pi(\lambda^{m},\omega^q) 
\end{equation}
contributes to the probability for events where all but $m$ gamma are in the 180-degree detector. To calculate the probabilities for all 180-degree coincidence events one considers all possibilities given by the sum on the right hand side of Eq.~\eqref{eqn:eqn25}. However, given the equality, the left hand side of that equation can be used to avoid the sum of permutations. Essentially, by using Eq.~\eqref{eqn:eqn25}, \textit{all possibilities where at least one gamma is in 180-degrees} is accounted for by a \textit{negation} from \textit{all possible events} with reference to the GOI. At this point, the reader is reminded of the task at hand; the goal is to find a gamma matrix where the elements of the matrix correspond to events where the gamma corresponding to that element is observed in 180 degree coincidence with any gamma. Therefore, the gamma matrix must use the matrices given in the left hand side of equation 25 to calculate the 180 degree correction matrix that is to be subtracted from the decay matrix given by $\Gamma(A,B)$. Given the multiple matrices present in equation 25, \textit{viz.}, a linear combination of matrices that provide the probabilities for 180 degree coincidences, one must be careful of the non-linearity of the SMF. This means when accounting for the linear combination of input matrices, the final probability matrix can be found by using a sum of multiple gamma matrices\footnote{Non-linear in the sense that the linear combination of matrices can not be inputted into the formalism and each term must be treated separately and then summed together}. In doing so, the summing-out correction can be given by the following:
\begin{equation}
    C_O = \Gamma(A,B) - \Gamma(A,\Lambda),
\end{equation}
or rather, recovering original notation.
\begin{equation}
    C_O = \Gamma(A,\Lambda_1) - \Gamma(A,\Lambda_2).
\end{equation}
\subsection{Summing-in}
Formulating the summing-in correction via the 180 degree method is much easier. Here, a sum of permutations does not need to be considered. Rather, only the events where a 180-degree coincidence occurs with a full energy gamma need to be considered. The summing-in terms contained in the A matrix remain; although, the sum starts at $k=2$ as only the product terms are required. Note, although $A$ was first derived to account for summing-in as a detection of the summed energy, the detection of the two gamma-rays that make up the summing being detected at 180-degree coincidence gives the same probability. However, to ensure the detection of the 180-degree coincidence event, the extra gamma-rays are required to not be in two detectors instead of just the one. Therefore, the 180 degree correction for summing-in is
\begin{equation}
    C_I = \Gamma(A-a,\Lambda_2).
\end{equation}
Using both the 180-degree summing-in and out correction, the decay matrix can be corrected as follows:
\begin{equation}
    S\rightarrow S^\prime = S +C_O - C_I.
\end{equation}
As is shown more explicitly in the following sections, this correction does not completely cancel out the summing effects. The \textit{true decay matrix}, i.e., the decay matrix for which no summing effects are present, would have the following form:
\begin{equation}
T = \mathcal{D}([f\chi]_i),
\end{equation}
where
\begin{equation}
    \chi = \sum_{k=0}^n x^k.
\end{equation}
Expanding out the branching terms gives
\begin{equation}
    T = \sum_{\mu = 1}^n\sum_{m=1}^\mu f_\mu\tau^{(\mu m)},
\end{equation}
where
\begin{equation}
    \tau_{ij}^{(\mu m)} = a_{ij}x_{\mu i}^{m-1}.
\end{equation}
In the following section, the correction in terms of multiplicity is given and the deviation of the 180-degree correction from the true decay matrix as a function of multiplicity is calculated.

\subsection{Multiplicity Expansion and Summing Inseparability}
To arrive at the multiplicity dependence, the branching ratios are extracted as done before:
\begin{equation}
    C_O = \sum_{\mu =1}^n f_\mu \left(\psi^{(\mu)} (A,B) -\psi^{(\mu)} (A,\Lambda) \right),
\end{equation}
\begin{equation}
    C_I = \sum_{\mu =1}^n f_\mu \psi^{(\mu)} (A-a,\Lambda). 
\end{equation}
Repeating the same steps as done previously in Eqs.~\eqref{eqn:eqn17}-~\eqref{eqn:eqn20}, the following is derived for the multiplicity expansion of the corrections:
\begin{equation}
    C_O = \sum_{\mu =1}^n \sum_{m=1}^{\mu}f_\mu \left(\phi ^{(\mu m) }(A,B) -\phi^{(\mu m)}(A,\Lambda)\right),
\end{equation}
\begin{equation}
    C_I = \sum_{\mu =1}^n \sum_{m=1}^{\mu}f_\mu\phi^{(\mu m)}(A-a,\Lambda),
\end{equation}
where,
\begin{equation}
    \phi_{ij}^{(\mu m)}(A-a,\Lambda) = \phi_{ij}^{(\mu m)}(A,\Lambda) - \phi_{ij}(^{(\mu m)}a,\Lambda),
\end{equation}
with,
\begin{equation}
    \phi_{ij}^{(\mu m)}(a,\Lambda) =a_{ij}\sum_{p=0}^ {m-1}\lambda_{\mu i}^p\lambda_{j0}^{m-p-1}.
\end{equation}
The deviation $\Delta$ of the correction to the complete correction, i.e., the correction that would transform $S$ into $T$, is the following.
\begin{equation}
\begin{split}
    \Delta &= T - 2\Gamma(A,B)  + \Gamma(A,\Lambda)+\Gamma(A-a,\Lambda)\\
    &=\sum _{\mu =1}^n\sum_{m=1}^\mu f_\mu \delta^{(\mu m)},\\
\end{split}
\end{equation}
where the multiplicity correction $\delta_{ij}^{\mu m}$ is defined as
\begin{equation}
    \begin{split}
        \delta_{ij}^{(\mu m)} =& \tau^{(\mu m)} -  2\phi^{(\mu m)}(A,B) + 2\phi^{(\mu m )}(A,\Lambda) - \phi^{(\mu m)}(a,\Lambda)\\
    =& a_{ij}x_{\mu i}^{m-1} -\sum_{p=0}^{m-1}a_{ij}\lambda_{\mu i}^p\lambda_{j0}^{m-p-1} - \\ & \sum_{l=1}^m \sum_{p=0}^{m-l}\frac{2a_{ij}^l}{N^{l-1}}\left[b_{\mu i}^p b_{j0}^{m-l-p} -\lambda_{\mu i}^p\lambda_{j0}^{m-l-p} \right].
    \end{split}
\end{equation}
A reminder on the notation here should be stated: the powers in the multiplicity expansion terms are matrix powers where the indices are taken \textit{after} the matrix products, i.e.,  $a^l_{ij} := [\boldsymbol{a^l}]_{ij}$. Equation 42 thus gives us the deviation between the 180-degree correction and the complete correction. The incompleteness of the 180-degree method comes from the inseparability of the 180-degree coincidence events and the coincident summing events. The equivalence between the summing events and the 180-degree coincidence events only holds exactly for $n=2$ decay schemes:
\begin{equation}
    \Delta = f_1\delta^{(11)} + f_2 (\delta^{(21)}+\delta^{(22)}),
\end{equation}
\begin{equation}
    \delta_{ij}^{(1 1)} =  a_{ij} \mathbb{1}_{1i}(1-\mathbb{1}_{j0}),
\end{equation}
\begin{equation}
    \delta_{ij}^{(2 1)} = a_{ij} \mathbb{1}_{2i}(1-\mathbb{1}_{j0}),
\end{equation}
\begin{equation}
    \delta_{ij}^{(2 2)} = a_{ij}(x_{2i} - \mathbb{1}_{2i}x_{j0} - \mathbb{1}_{j0}x_{2i}),
\end{equation} 
For $\delta_{ij}^{(11)}$, there are no values for $i$ and $j$ that make it nonzero given the conditions in the identity matrices present. The $\mu=2$ terms separately equal zero for all but the $i,j=2,1$ term, which cancels out with $\delta_{21}^{(21)}$ added to $\delta_{21}^{(22)}$ with the normalization condition of the transition matrices realized:
\begin{equation}
    \delta_{21}^{(21)} + \delta_{21}^{(22)} = a_{21} (1-x_{10}).
\end{equation}
By defining the set of all summing events $S$ and the set of all 180 coincident events $C$, these sets are inseparable and the intersection of these sets is what defines the deviation derived above. The deviation thus defines the statistical bound for the 180-degree correction method. If the 180-degree method is used to sufficiently measure the coincidence summing, it is only statistically sufficient within this bound, a bound that is dependent on ontic multiplicity and the number of detectors involved, with a greater number of detectors decreasing the deviation between the summing events and the 180 degree coincident events. When using this method to correct for coincidence summing, this deviation must be calculated and compared to the statistical uncertainty of the peak data to check if it contributes considerably to the overall uncertainty of the measurement at hand.

\subsection{Predictions for $\Delta$ with toy model decay.}
\begin{figure}[]
    \centering
    \begin{subfigure}[t!]{\columnwidth}
        \centering
        \includegraphics[width=.95\textwidth]{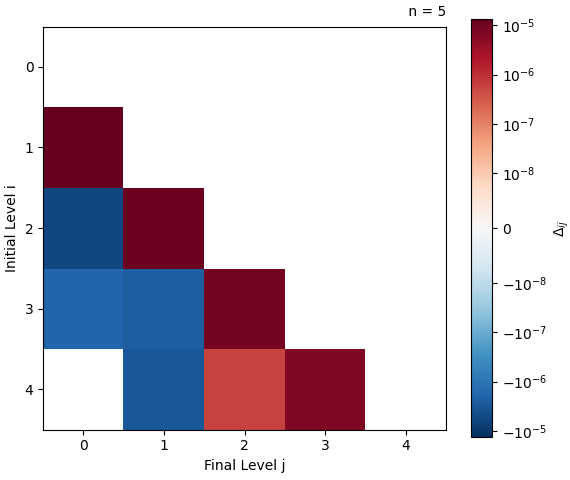}
        
    \end{subfigure}
    
    \begin{subfigure}[c!]{\columnwidth}
        \centering
        \includegraphics[width=.95\textwidth]{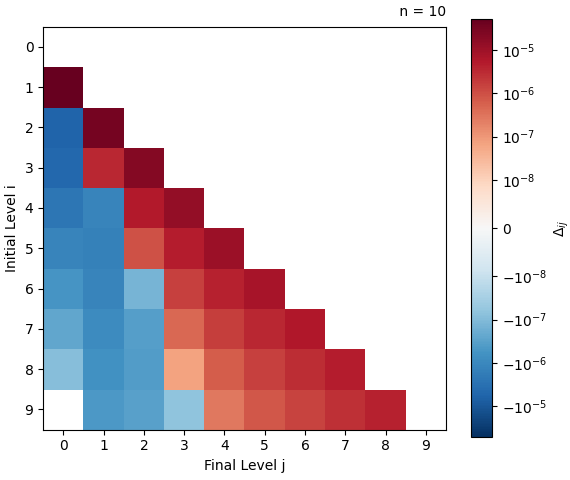}
        
    \end{subfigure}
    
    \begin{subfigure}[c!]{\columnwidth}
        \centering
        \includegraphics[width=.95\textwidth]{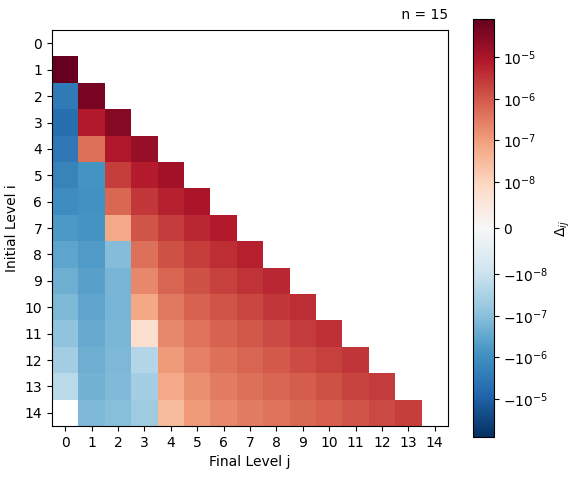}
        
    \end{subfigure}
    
    \caption{Delta matrix values for n = 5, 10, and 15 for the toy model decay scheme. The plots take the form of the transition matrices where each square indicates an element of the matrix and thus a transition in the decay scheme.}
    \label{fig:fi6}
\end{figure}
Values for $\Delta$ can be estimated for different multiplicities by using a toy model decay scheme. By using a decay scheme of equal energy spacing and equal transition probabilities per level, the number of levels can be scaled and the $\Delta$ matrix can be calculated to show the behavior of the 180-degree corrections at higher order across the different transitions in the decay. The elements of the branching vector are then $f_i = 1/n$ and the elements of the transitions matrix are $x_{ij} = 1/i $ for $ i > 1$. The energy spacing for each level is $100$ keV with all transitions allowed; thus the energy of the transition from the top level to the ground state is $n\times100$ keV. The efficiencies for these energies are taken as simulated efficiencies for GRIFFIN.
With this decay scheme, the $\Delta$ matrix values are shown in Fig.~\ref{fig:fi6} for decays with $n=5,10,15$ levels. All three plots have similar orders of magnitude for the range for their deviations, ranging from negative $10^{-5}$ to positive $10^{-5}$; that is, from negative $10^{-3}\%$ to positive $10^{-3}\%$. The inflection from negative to positive deviation marks the crossover where the summing-in and summing-out corrections contribute equally. The transitions with the negative deviations are largely affected by summing-in as they are transitions over a larger number of levels. As you approach the diagonal, the transitions approach single level transitions where no summing-in occurs. However, for these transitions summing-out is largely prominent as they may occur in cascades with a large number of gamma-rays. As 180-degree coincidence events are still corrupted by summing, the corrected $S$ matrix will still be larger than $T$ for the summing-in dominant transitions and smaller then $T$ for the summing-out dominant transitions. The deviation is also stronger for transitions from levels closer to the ground state as those transitions are more likely to occur given the branch feeding from above. 

\subsection{Coincidence Summing In the decay of $^{133}$Ba} %
\begin{table}[h!] 
    \centering
    \begin{tabular}{||c|c|c|c|c||}
       \hline
$I\epsilon$& $E_\gamma$ (keV) & $x_\gamma$& $\epsilon_\gamma$ &$\Delta_\gamma(\times 10^{-6}$)\\
\hline
$< 0.002$&81.0 & 1.000  & 0.456 &  9.69\\  
\hline
$<0.001$&160.6 & 0.194 & 0.401& -5.08\\
&79.6 & 0.806 &0.4565& 7.84\\
\hline
0.145& 383.8 & 0.322 & 0.239 &  -0.91\\
&302.8 & 0.661 &  0.343& 1.28\\
&223.2 & 0.0163 &  0.283 & 0.20\\
\hline
0.855& 356.0 & 0.870&  0.253 &  -4.11\\
&276.4  & 0.100 &  0.301&  4.93\\
&53.2  & 0.300 &  0.440&  9.80\\  
\hline
    \end{tabular}
    \caption{Decay and simulated GRIFFIN detection probabilities for the gamma-rays emitted in the EC decay of $^{133}$Ba. The third column, $x_{\gamma}$, are the transition probabilities for each gamma transition. The forth column, $\epsilon_\gamma$, are simulated efficiencies for GRIFFIN (11 cm, Single crystals, no shields, US+DS SCEPTAR, no delrin)\cite{Mills2015}. The final column, $\Delta_\gamma$, contains the deviation between $T_\gamma$ and $S_\gamma$ after applying the 180-degree summing corrections: $S^\prime \rightarrow S +C_O - C_I$. All $^{133}$Ba decay data is collected from ENSDF\cite{ENSDF_Ba133}.}
    \label{tab:placeholder}
\end{table}

The EC decay of $^{133}$Ba to $^{133}$Cs serves as a simple example to study coincidence summing-in detail as the $Q$-value being rather low limits the number of available states making the transition matrix manageable. The decay and simulated GRIFFIN detection probabilities are shown in Table 1. The simulated GRIFFIN data used are the simulated efficiencies calculated from the available online efficiency calculator \cite{Mills2015}. For each energy, the columns show, in order: the transition probability, calculated from relative intensities; the simulated GRIFFIN efficiency at that energy; and the deviation between the probabilities of detection without and with coincidence summing, after applying the 180-degree corrections. The deviation between the true and corrected probabilities are on the order of $10^{-4}\%$ and are comparable to the toy model predictions. The transitions that pass over multiple levels have a negative deviation as expected from the results of the toy model due to the higher likelihood of summing-in; these can be seen in the $160.6$ keV, $383.8$ keV, and the $356.0$ keV transitions. The $81.0$ keV and the $53.2$ keV transitions have a larger deviation due to their increased likelihood of summing-out; for high multiplicity events, the 180-degree coincidences still experience summing-out and thus the correction is still limited. By allowing any combination of gamma-rays to be in the 180 detector to the GOI, higher-order summing \textit{is} accounted for to some degree but the GOI is still required to be detected and remaining gamma-rays can still cause summing-out.

The EC-decay of $^{133}$Ba serves as a good example to see the lower limits of this deviation where the 180-degree correction holds to a sufficient precision. For most decay spectroscopy experiments, deviations to photo-peak areas on the order of $10^{-6}$ are below the statistical uncertainty granted by the experiment. However, there are areas of study where such precision may be important. In high precision measurements of super-allowed beta decay used for precision test of the CKM anomaly, such orders of magnitude may start to take affect. In recent reviews of superallowed $\beta$-decay and CKM unitarity test, precision are at the level of $10^{-4}$ \cite{Gorchtein_2024,PhysRevC.102.045501} \footnote{Among the highest-precision superallowed $\beta$-decay measurements, there remain cases with comparatively larger uncertainties. Such cases motivate new experimental efforts, such as those underway with GRIFFIN at TRIUMF}. The $10^{-4}$ level is still above the limit where these deviations exist but with these corrections potentially being a systematic effect that might accumulate, these deviations may require attention as measurements increase in precision--especially with the GRIFFIN collaboration involved in the superallowed $\beta$-decay research program.

\begin{figure}[h]
    \centering
    \includegraphics[width=.9\linewidth]{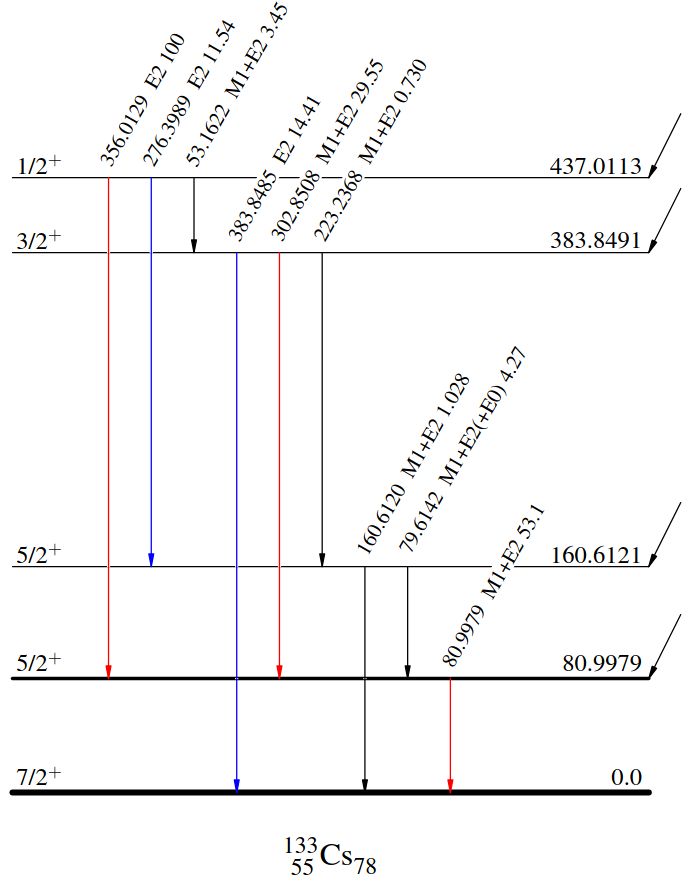}
    \caption{Gamma decay levels scheme from the electron capture of $^{133}$Ba. Figure from ENSDF \cite{ENSDF_Ba133}.}
    \label{fig:placeholder}
\end{figure}

\section{Gated Probabilities}
We now look to solving for the probabilities of \textit{gated} gamma-rays, i.e., those in coincidence with a second GOI in a second detector. Gates are often used in gamma spectroscopy to select from certain decay paths, calculate branching ratios, and calculate efficiencies. The formalism is extended to gated probabilities as follows. The first difference that must be accounted for here is the requirement of two gamma-rays needing to be specified. For the singles, one could easily address the single gamma of choice with the indices of the matrix, for the gated gamma-rays this would lead to having to derive a rank 4 tensor where all possible gates and observed gamma probabilities are given. To avoid this most likely unwieldy pursuit, the general gate is selected before hand and then the probability matrix for that gate is calculated. Again, this process starts by deriving some initial probabilities that are functions of the efficiencies:

\begin{equation}
    G = A \left(\frac{N-1}{N}\right).
\end{equation}
The above is the probability for $E_{ij}$ to be detected (by either the true gamma or summing-in) in any detector besides the detector that contains the gamma $\gamma_{gate}$ that is gated on. When deriving the probability matrices for the gated gamma-rays the entire matrix will have a factor of the probability for full detection of $\gamma_{gate}$. For a given gate $\gamma_{kl}$ the gamma-rays above and below this transition may be separated. The probability matrix that would be given for all gamma-rays in coincidence with this gate gamma is a partitioned matrix with the non zero elements making up two block matrices centered on the diagonal and separated by the indices of the gate. For example, for $n=5$ with a gate on $\gamma_{32}$ or on $\gamma_{41}$, the probability matrices would have the following form:
\begin{equation}
    Y^{(3,2)} = 
\begin{pmatrix}
 0 &\\
        y_{10} & 0 \\
        y_{20} & y_{21} & 0\\
0&0&0&0&  \\
       0&0&0& y_{43}&0 & \\
          0&0&0&y_{53}&y_{54}&0 \\
\end{pmatrix},
\end{equation}
\begin{equation}
    Y^{(4,1)} = 
\begin{pmatrix}
0&\\
y_{10}&0\\
0&0&0\\
0&0&0&0\\
0&0&0&0&0&\\
0&0&0&0&y_{54}&0
\end{pmatrix}.
\end{equation}
One can therefore separate the gated  probability matrix into a direct sum of strictly square lower triangular matrices (and the zero matrix $0$ if the indices for the gate are not consecutive) where the subscript denotes their dimension.
\begin{equation}
    Y^{(k,l)} = B^{\uparrow}_{(l+1)} \oplus 0_{(k-l-1)}\oplus B^{\downarrow}_{(n-k+1)}
\end{equation}
For $Y^{(4,1)}$,
\begin{equation}
    B^{\uparrow}  = \begin{pmatrix}
        0&\\
y_{10}&0\\
    \end{pmatrix},
\end{equation}
\begin{equation}
    0  = \begin{pmatrix}
        0&\\
0&0
    \end{pmatrix},
\end{equation}
\begin{equation}
    B^{\downarrow}  = \begin{pmatrix}
        0&\\
x_{54}&0\\
    \end{pmatrix},
\end{equation}
where $B^\uparrow$ gives the gamma-rays that are \textit{gated from above} and $B^\downarrow$ gives the gamma-rays that are \textit{gated from  below.} When considering all possible probabilities for the gated gamma-rays, the gated from above and the gated from below gamma-rays can be treated separately. This is because from a given gate all possible decay paths must go through that gated transition. In such a case, when considering the possible events that involve detecting a gamma below the gate, i.e., gamma-rays that are gated from above, all the gamma-rays that are above the gate, if any, will have probabilities that are not coupled to those below. Therefore, the probability factors present in $B^\uparrow$ that belong to gamma-rays above the gate can all be factored out as scalars that multiply onto the matrix with elements whose terms are distinct to the gammas below\footnote{To assist in the possibly confusing language; gamma-rays below the gate are gamma-rays that are gated from above; gamma-rays above the gate are gamma-rays that are gated from below}. These factors, denoted $h_\downarrow$ and $h_\uparrow$, are separated as follows:
\begin{equation}
    B^\uparrow = h_\downarrow \Gamma^\uparrow,
\end{equation}
\begin{equation}
    B^\downarrow = h_\uparrow \Gamma^\downarrow.
\end{equation}
Here, the SMF is extended and the up and down gamma functions $\Gamma^\uparrow$ and $\Gamma^\downarrow$ are defined as,
\begin{equation}
    \Gamma^{\uparrow(k,l)}(Q,\Lambda_\nu) = \mathcal{D}([\hat{l}\Lambda_\nu^\uparrow]_i)Q^\uparrow \mathcal{D}([\Lambda_\nu^\uparrow]_{i0}),
\end{equation}
\begin{equation}
    \Gamma^{\downarrow(k,l)}(Q,\Lambda_\nu) = \mathcal{D}([f\Lambda_\nu^\downarrow]_i)Q^\downarrow \mathcal{D}([\Lambda_\nu^\downarrow]_{i0}),
\end{equation}
where $f^\downarrow = (f_k,f_{k+1}...f_n)$ and $\hat{l}$ is denoted as the vector with dimension $l+1$ and only non-zero component being equal to one in the $l^{th}$ component. For example, with the gate on $\gamma_{32}$, $\hat{l} = (0,0,1)$. This vector will multiply the partitioned $\Lambda^\uparrow$ and then make up the diagonal matrix that multiplies $Q^\uparrow$ from the left. The separability of the gamma-rays above and gamma-rays below will allow us to partition the $\Lambda$ and $G$ matrices to calculate the factored block matrices $\Gamma^\uparrow$ and $\Gamma^\downarrow$. For the gamma-rays gated from above all branching comes from the $\gamma_{gate}$ branch or above it and thus factors out. Therefore, all branching terms in $B^\uparrow$ are present in $h^\downarrow$. When calculating the probabilities for $\Gamma^\uparrow$, all the gamma-rays are coming from the level in which $\gamma_{gate}$ transitions to--the level which gives the form for $\hat{l}$. For the $Q$ matrix here, $G$ is used as defined previously.
The elements of $G$ act as the full energy gamma-rays of interest but weighted by the probability $(N-1)/N$, of not being in the detector with $\gamma_{gate}$. To account for summing-out, the $\Lambda_2$ matrix is used from the 180-degree corrections as the extra gamma-rays are to not be in the two detectors required for the coincidence:
\begin{equation}
    \Gamma^\uparrow(k,l) = \mathcal{D}([\hat{l}\Lambda_2^\uparrow]_i)G^\uparrow \mathcal{D}([\Lambda^\uparrow_2]_{i0}),
\end{equation}
\begin{equation}
   \Gamma^\downarrow(k,l)  = \mathcal{D}([f^\downarrow\Lambda_2^\downarrow]_i)G^\downarrow \mathcal{D}([\Lambda_2^\downarrow]_{i0}).
\end{equation}
 The factors $h^\uparrow$ and $h^\downarrow$ from equations 49 and 50 come from the summing above and below the gate, respectively. In the SMF, the left multiplying diagonal matrix accounts for the summing-out from gamma-rays above a GOI while the right multiplying diagonal matrix accounts for the summing-out below a gamma. The elements of these diagonal matrices that are calculated from the whole summing-out matrix and branching vector, i.e., not the partitioned ones, can be used to calculate these factors:
\begin{equation}
    B^\uparrow(k,l) = [f\Lambda_2]_{k}\Gamma^\uparrow(k,l) ,
\end{equation}
\begin{equation}
    B^\downarrow(k,l) = [\Lambda_2]_{l0} \Gamma^\downarrow(k,l).
\end{equation}
Our final gated probability matrix is thus:
\begin{equation}
    Y^{(k,l)} = A_{kl}Z^{(k,l)}(G,\Lambda_2),
\end{equation}
where
\begin{align}
    Z^{(k,l)}(G,\Lambda_2) =& [f\Lambda_2]_{k}\Gamma^{\uparrow(k,l)}(G,\Lambda_2) \oplus 0_{(k-l-1)}\oplus\\\nonumber
    & [\Lambda_2]_{l0} \Gamma^{\downarrow(k,l)}(G,\Lambda_2) .
\end{align}
 and $A_{kl}$ is the probability for the gate where summing-in can occur. This method of calculating gated probabilities via equation 55 is the \textit{partitioned matrix formalism} (PMF)\footnote{We avoid calculating the multiplicity expansion for the gated probabilities due to the complexity of the expression}. Using this formalism, the form of the 180-degree gated coincidences can be given.
\subsection{Summing Corrections}
\subsubsection{Summing-out}
When attempting to sufficiently measure the gated summing probabilities, the summing with both the $\gamma_{gate}$ and the GOI must be considered--separately and then simultaneously. In practice, this would look like collecting multiple spectra: One spectrum for $\gamma_{gate}$ and GOI in coincidence--with a 180-degree coincidence with the GOI; one for $\gamma_{gate}$ and GOI in coincidence--with a 180-degree coincidence with $\gamma_{gate}$; another where both $\gamma_{gate}$ and the GOI have a 180; and finally, the summing of $\gamma_{gate}$ and the GOI must be accounted for, so a 180-degree coincidence of $\gamma_{gate}$ and GOI is needed. The first two account for when only one summing event with an extra gamma-ray occurs, and the third accounts for when two summing events with an extra gamma-rays occur. The zeroth-order gated coincidence is considered as the summing of $\gamma_{gate}$ and GOI; first-order as the summing of $\gamma_{gate}$ and extra or GOI and extra; and second-order gated coincidence summing as the summing of gate with extra and gamma with extra. Following the same process as before, first order gate coincidence summing can be given by 
\begin{equation}
    \begin{split}
         \sum_{k=1}^n\lambda_{2}^k& =\sum_{k=1}^n\sum^k_{q=0} \pi(\lambda_{3}^{k-q},\omega^q)\\
         \Lambda_2 - \Lambda_{3}  & = \Omega + \sum_{m=1}^{n-1}\sum_{q=1}^{n-m}\pi(\lambda_{3}^{m-q},\omega^q) \\
           \Lambda_2 - \Lambda_{3}  & = \Omega + \sum_{m=1}^{n-1}
         \Pi_m,
    \end{split}
    \label{eqn:eqn65}
\end{equation}
where 
\begin{equation}
     \Pi_m = \sum_{q=1}^{n-m}\pi(\lambda_{3}^{m-q},\omega^q).
\end{equation}
The left hand side of Eq.~\eqref{eqn:eqn65} thus gives the permutations of all extra gamma-rays where at least one is in a selected detector. No angular correlations have been included, so by using the probability of hitting a single detector, the 180-degree detector of the GOI is selected. Moreover, this fact can be continually used to account for both scenarios of the first order gated coincidence summing by multiplying the probabilities by two. The above matrices and the PMF with the non-linearity condition are used to arrive at the first order correction for the summing-out of the gated probabilities:
\begin{equation}
    Y_{C_O}^{(k,l)_{(1)}} =2A_{kl}\left[Z^{(k,l)}(G,\Lambda_2) - Z^{(k,l)}(G,\Lambda_3)\right].
\end{equation}
For the second order correction, at least two gamma-rays are required to be in certain detectors, with the extra gamma-rays not in the four detectors used for the quad-coincidences. Building on this idea, the following correction matrices are given, where the permutations with $\omega$ terms of $q=1$ are negated as at least $q=2$ is required:
\begin{equation} 
    \begin{split}
         \sum_{k=1}^n\lambda_{3}^k& =\sum_{k=1}^n\sum^k_{q=0} \pi(\lambda_{4}^{k-q},\omega^q)\\
         \Lambda_3 - \Lambda_{4}  & = \Omega + \sum_{m=1}^{n-1}\sum_{q=1}^{n-m}\pi(\lambda_{4}^{m-q},\omega^q) \\
          \Lambda_3 - \Lambda_{4} - \sum_{k=1}^n\pi(\lambda_{4}^{k-1},\omega) & = \Omega + \sum_{m=1}^{n-1}\sum_{q=2}^{n-m}\pi(\lambda_{4}^{m-q},\omega^q) \\
           \Lambda_3 - \Lambda_{4} - \mathcal{P}& = \Omega + \sum_{m=1}^{n-1}\sum_{q=2}^{n-m}\pi(\lambda_{4}^{m-q},\omega^q), \\
    \end{split}
\end{equation}
where
\begin{equation}
\mathcal{P} = \sum_{k=1}^n\pi(\lambda_{4}^{k-1},\omega).
\end{equation}
The second order corrections would then be
\begin{equation}
    Y_{C_O}^{(k,l)_{(2)}} =A_{kl}\left[Z^{(k,l)}(G,\Lambda_3) - Z^{(k,l)}(G,\Lambda_4)- Z^{(k,l)}(G,\mathcal{P})\right].
\end{equation}
For the majority of realistic decays, the second-order correction will be small if not zero, given the need for at least four gamma-rays. However, when looking at $\beta^+$-decay where back-to-back 511 keV emission can occur from positron annihilation, this second-order term may be larger. For the zeroth-order term, the GOI is required to be in 180-degree with $\gamma_{gate}$ and the extra gamma-rays are not in two detectors. To achieve this, the $\Omega$ and $\Lambda_2$ matrices are used:
\begin{equation}
    Y_{C_O}^{(k,l)_{(0)}} =A_{kl}Z^{(k,l)}(\Omega^P,\Lambda_2),
\end{equation}
where $\Omega^P$ contains the peak efficiencies, rather than total efficiencies; a necessary change due to the need to have the peak energy gamma in the opposite detector, not just any energy.
\subsubsection{Summing-in} 
To get the summing-in, one must attend to the summing-in for $\gamma_{gate}$ and summing-in for the GOI. For summing-in for the GOI, the same notion that is done in the singles can be followed; the first term of the sum of gated probabilities is subtracted off and the summing-out $\nu$ value increases by one. 
\begin{equation}
    \text{Summing-into gamma term: } A_{kl}Z^{(k,l)}(G-g,\Lambda_3),
\end{equation}
\begin{equation}
    \text{Summing-into gate term: } (A-a)_{kl}Z^{(k,l)}(G,\Lambda_3.)
\end{equation}
These cases need to be considered separately and simultaneously, for a total of 3 terms. With each case considered separately, the first order summing-in terms are given; when considered simultaneously, the second order terms are given:
\begin{align}
    \nonumber
    Y^{(k,l)}_{C_I} = &A_{kl}Z^{(k,l)}(G-g,\Lambda_3)+(A-a)_{kl}Z^{(k,l)}(G,\Lambda_3)+ \\
    &(A-a)_{kl}Z^{(k,l)}(G-g,\Lambda_3).
\end{align}
\subsubsection{Summing Inseparability}
Due to the complexity of the gated probabilities, the multiplicity expansion that affords the desired truncation of the terms is avoided. However, the form of the deviation between the gated 180-degree correction and the true correction is still provided. The \textit{true gated matrix} $T_g$, whose form is analogous to the singles, is given by
\begin{equation}
    T_g = a_{kl}\left[[f\chi]_{k}\mathcal{D}([\hat{l}\chi^\uparrow]_i)g^\uparrow\oplus 0_{(k-l-1)}\oplus \mathcal{D}([f^\downarrow\chi^\downarrow]_i)g^\downarrow \right].
\end{equation}
Subtracting off the gated probabilities, the summing-out corrections, and adding the summing-in corrections (both to full order):
\begin{equation}
    \Delta_g = T_g - Y^{(k,l)} - Y^{(k,l)}_{C_O} +Y^{(k,l)}_{C_I}.
\end{equation}

The magnitude of this deviation largely depends on the gate chosen. If the transition being gated on skips over large parts of the decay scheme, the multiplicity for a decay with that gate may be much lower than the maximum multiplicity defined by the decay scheme. However, with the extra detectors required for the coincidence measurements, the likelihood of losing a coincidence event to summing increases. The determination of large levels schemes via decay spectroscopy resides heavily on the ability to gate on certain transitions to calculate transition probabilities and angular correlations. The PMF may aid in handling the combinatorics of different coincidences when analyzing measurements for nuclear structure. Indeed, it shows--via equations 61 and 62--that when calculating probabilities for transitions on one side of a gate, the transitions/branching probabilities on the other side of the gate factor out. This fact is crucial in determining relative intensities of transitions originating from a single level; that is, when taken in ratio, feeding probabilities from across the gate cancel--as well as do summing-out affects. More work needs to be done in investigating coincidence probabilities. In particular, an incorporation of angular correlations into the formalism is much needed to arrive a more accurate model of coincidence probabilities.

\section{Remarks and Conclusion}
\subsection{Compton Summing and Compton Coincidence}
When investigating summing-in and gated probabilities, only true photo-peak summing was considered. When calculating summing-out, Compton-scattered probabilities are included rather easily as the summing effect does not depend on the energy the gamma scatters to. However, in gamma-decay measurements, it is possible for summing in to occur between one of the gamma-rays of the sub transition (that is, if A and B can sum to C, than A and B are sub transitions of C) and any other gamma of high enough energy that by chance Compton-scatters to just the right energy to cause a count in the photopeak of the summed energy transition. 
\par
For a general decay scheme, calculating the probabilities for Compton summing-in is rather non-trivial as it requires an energy relation between the possible true summing-in gamma-rays and all other higher energy gamma-rays above or below. For example, in Figure 5, when considering the summing-in to the $x_{20}$ transition--normally occurring via the $x_{21}$ and $x_{10}$ summing, if $x_{32}$ happens to be at an energy greater than $x_{21}$ and $x_{10}$, then a Compton-scattering of the $x_{32}$ could sum with either transition to give an energy of $x_{20}$. 
\par
In practice, however, the 180-degree correction deals nicely with this concern as any possible higher energy transitions that have Compton-scattered into the right energy are given in the photo-peak. Notice here, that is not simply an event in the smooth Compton background. Coincidence events where one of the gamma-rays is a full energy gamma-ray and the other is a Compton-scattered gamma-ray still contribute a peak to coincidence spectra. This can be seen when taking a transition to be in coincidence with itself: if a true coincidence is not possible, a photopeak still appears in the spectrum from the true gamma-ray and the gamma-ray produced from the Compton-scattering of a higher energy gamma-ray. Nonetheless, the deviations presented in this paper still provide an understanding on the precision of the 180-degree correction method for true coincidence summing.
\par
Similar to Compton summing-in, which occurs in a single detector, the same coincidence can occur with the gamma-rays in separate detectors--which may inflate the gated peaks discussed in the second half of this paper. The difficulty with generalizing these events in the coincidence case still stands. However, both of these effects, of course, are relatively weak when compared to the photopeak summing-in and coincidences--though still important for most gamma spectroscopy measurements.
\subsection{Conclusion}
The formalism of Semkow \textit{et al.} is generalized to include selective coincidence events and extended into a multiplicity expansion, the terms of which contribute increasingly towards the likelihood of a summed event. Using the SMF, the probabilities for 180-degree coincidence events as a method to sufficiently measure the coincidence summing is provided. This method is shown to not be completely sufficient such that there is a nonzero separability between the summed events and the 180-degree coincidence events that results in a deviation from a complete correction. This deviation increases with ontic multiplicity of the event. When using this method to correct for coincidence summing, the resultant deviation needs to be calculated and compared to the statistical precision of the experiment. A toy model analysis of this deviation is given and shows where in a general decay scheme summing-in and summing-out  effects still dominate after correction. Using simulated peak efficiencies of the GRIFFIN spectrometer, these deviations are shown to be on the order of $10^{-5}$ at most--far below the limit for most decay spectroscopy experiments, though might accumulate for high-precision measurements of superallowed beta decays. 
\par
The probabilities for gated gamma-rays that include coincidence summing effects are calculated using the partitioned matrix formalism which further extends the SMF; the 180-degree coincidence corrections for these gated probabilities are provided. For the development of complicated nuclear level schemes, gated energy spectra are often used to deduce particular transition probabilities of certain cascades or to compute angular correlations to assign spin of excited states. Understanding the combinatorics of all the different coincidence summing in these gated spectra is crucial. The PMF allows one to handle the combinatorics of this process by providing the different summing coincidences.  The extension to gated probabilities is complicated by the order of summing present, such that with two gamma-rays involved, different degrees of summing can occur. The 180-degree summing correction probabilities for zeroth-, first-, and second-degree summing are given along with deviation from complete correction. The goal of this work was ultimately to check the limit of the 180-degree coincidence method as a method to sufficiently measure coincidence summing and to provide a way to calculate coincidence summing probabilities for gated spectra. The GRIFFIN collaboration at TRIUMF focuses heavily on high precision superallowed $\beta$-decay measurements and investigations into nuclear structure. To this end, knowledge on the limits of this correction techniques is pertinent.  This work has showed that though not exact, this method still holds.

\section*{Acknowledgments}
The author would like the thank Konstantin Stoychev and Vinzenz Bildstein of the University of Guelph for their helpful review of this manuscript. This work was supported by the Natural Sciences and Engineering Research Council of Canada (NSERC).
\bibliographystyle{elsarticle-num} 
\bibliography{biblio}

\end{document}